# Penetration of alkali atoms throughout graphene membrane: theoretical modeling

D. W. Boukhvalov[1] and C. Virojanadara[2]

[1] *School of Computational Sciences, Korea Institute for Advanced Study (KIAS) Hoegiro 87, Dongdaemun-Gu, Seoul, 130-722, Korean Republic*

[2] *Department of Physics, Chemistry, and Biology, Linköping University, S-581 83 Linköping, Sweden*

*Theoretical studies of penetration of various alkali atoms (Li, Na, Rb, Cs) throughout graphene membrane grown on silicon carbide substrate are reported and compared with recent experimental results. Results of first principles modeling demonstrate rather low (about 0.8 eV) energy barrier for the formation of temporary defects in carbon layer required for the penetration of Li at high concentration of adatoms, higher (about 2 eV) barrier for Na, and barriers above 4 eV for Rb and Cs. Experiments prove migration of lithium adatoms from graphene surface to the buffer layer and SiC substrate at room temperature, sodium at 100°C and impenetrability of graphene membrane for Rb and Cs. Differences between epitaxial and free standing graphene for the penetration of alkali ions are also discussed.*

E-mail: danil@kias.re.kr; chavi@ifm.liu.se

## 1. Introduction

Graphene is the novel and promising material for the various applications.[1-4] Possible application of graphene in chemical sensor,[4-9] composite materials and supercapasitors,[10-14] and metal-free catalyst[15-18] requires detailed study of the interaction of various adsorbates with graphene and especially litium.[19] Recent experimental[20] and theoretical[21] studies suggest for the impenetrability of graphene membrane for the helium and hydrogen atoms. Only large vacancies with the size above 5 Å (two lattice parameters) suggested by theory,[21-23] and experiment[24] for the possible penetrable pores in graphene. However, our recent experimental studies of functionalization of epitaxial monolayer graphene, prepared by sublimation of SiC crystal, using hydrogen and alkaline metals show that is not always the case.[25-29]

Epitaxial graphene preparation by sublimation of a SiC wafer at high temperature in Ar at atmospheric pressure is known to result in homogeneous monolayer graphene.[28,30,31] However the monolayer graphene prepared this way have an unwanted extra layer located in between the graphene sheet and the SiC substrate. This layer the so called buffer layer[32-34] is the first carbon layer that has strong covalent bonding with the SiC substrate and has no graphitic properties. Therefore, efforts to decoupling this buffer layer from the SiC substrate were intensively studied. Experimental investigation on functionalization of monolayer graphene grown on 6H-SiC(0001) substrates using hydrogen and alkaline metals were performed.[25-29] The hydrogen was found to go through the graphene and the buffer layer at a substrate temperature of 700°C and successfully decouple the covalent bonding at the interface region by passivating the outermost Si atom dangling bonds.[25,28] This in turn transforms the carbon buffer layer to a graphene sheet. The mechanism of the hydrogenation process suggested from STM measurements[29] is that the hydrogen initiate some defect on the graphene sheet and penetrates through the existing carbon layers and consequently passivates the Si atoms at the interface. This was concluded from the island like feature observed to form randomly on the surface after initial atomic hydrogen exposures. These islands were found to coalesce together and become larger in size with larger exposures. A similar results was also observed after Li deposition[26,27] but the defects were much more visible than the ones in the hydrogen case,[25] see Fig. 1. Moreover, the Li was reported to reach the interface region directly after deposition at room temperature.[26,27] Experimental studies of other alkaline metals, Na, Rb and Cs, were also recently reported.[35,36] For Na a low temperature annealing was found to be required for the Na to penetrate to the interface region.[36] On the other hand no penetration of Rb and Cs metals was observed at RT or even after annealing up to the metal's melting points.[35] See refs. [26, 27, 35, 36] for experimental details.

In order to gain more understanding of the intercalation/ penetration mechanism of adsorbates on the epitaxial graphene sheet, a set of alkaline metals was selected and the detailed theoretical studies were performed and presented in comparison to the recent experimental findings.[26,27,35,36] Moreover, discussed controversy of the results for alkaline metals penetration throughout carbon layers with previous theoretical results[20, 21] is also of interest. Main difference among alkali elements and helium, hydrogen is the significant doping of the graphene.[37-42] Previous theoretical works suggest for the essential changes of the chemical properties[43] and vacancies formation energetics[44] in doped graphene. For the modeling of the interaction of the epitaxial graphene on silicon carbide substrate with alkali atoms, a minimal model of epitaxial graphene feasible for the description of its chemical properties[45] have been used.

## 2. Computational method and model

The modeling was carried out by density functional theory (DFT) realized in the pseudopotential code SIESTA,[46] as was done in our previous work.[45] For the hydrogenation processes all calculations were done using local density approximation (LDA)[47] which is the most suitable for the description of epitaxial graphene chemical properties.[45] Full optimization of all atomic positions and lattice parameters was performed. All calculations have been done with using spin-polarized mode. During the optimization, the electronic ground state was found self-consistently using norm-conserving pseudo-potentials for cores and a double-ζ plus polarization basis of localized orbitals for carbon and alkali atoms, and double-ζ basis for hydrogen. Optimization of the forces and total energies was performed with an accuracy of 0.04 eV/Å and 1 meV, respectively. All calculations were carried out for an energy mesh cut off of 360 Ry and a k-point mesh 8×8×4 in the Mokhorst-Park scheme.[48] Used model of epitaxial graphene contain two carbon layers with adsorbed alkali atoms from top and hydrogen atoms from bottom and not strictly a two-dimensional system. Adsorption of chemical species leads to additional deviation from the planar geometry. Therefore, we have chosen several *k*-points (namely, four) also in the *z* direction.

For our modeling of the interaction of graphene on SiC substrate have been used the minimal model of epitaxial graphene (see Fig. 2). This model was used before[24] for the explanation of the special chemical properties of epitaxial graphene. This model contains two layer of graphene. Upper layer correspond with graphene layer and lower layer partially hydrogenated with coverage 12.5% (4 hydrogen atoms for 32 carbon atoms in supercell) correspond with buffer layer formed between graphene and SiC substrate. The distance between described bilayers along *z* direction is 20Å that provides perfect separation of graphene layers. For comparison the calculation for the penetration of adatoms throughout free standing graphene monolayer was also performed. We performed a modeling of the penetration of alkali atoms (Li, Na and Rb, Cs) used in experiment at different coverages. 100% coverage is corresponding with the experimentally observed situation of the Li droplets on the graphene (see Fig. 1) and 12.5% coverage corresponds with the single alkali adatom on graphene. For calculation of the energy barriers of the process of penetration on alkali atom the position of this atom and one of the nearest carbon atoms have been changed. Full optimization of atomic position and lattice parameters with fixed positions of alkali atom in substitutional position and two most remote carbon atoms have been performed (see Fig. 2b and e). Obtained atomic structure is close to the atomic structure for the moment of the situating of alkali atom on the level of graphene membrane. The value of energy barrier has been calculated by standard formula: $E = (E_{N+1} - E_N)/n$, where $E_N$ is the total energy of the system at the $N^{th}$ steps of penetration (odd steps correspond to the substitution of carbon atoms by alkali (Fig. 2 b and e), even steps to situation after penetration of $N/2^{th}$ atom to the other side of graphene membrane (Fig. 2 c and f)), and n is the total number of alkali adatoms on for the studied type of coverage. Division of the energy difference for the number of impurity atoms require for the minimize influence of the choice of the size of supercell and type of coverage and permit estimate the energy required for the penetration of moll of alkali adatoms.

## 3. Results and discussions

Recent experimental results of Li intercalation on epitaxial graphene growth on SiC(0001),[26-27] showed that Li could intercalate directly after deposition at room temperature and transform the buffer layer to an additional graphene sheet. This suggestion was concluded from the electron reflectivity curves and high resolution photoemission experiments. After deposition, Li was detected in between the carbon layers and at the interface between the buffer layer and the SiC substrate[26] where the latter, a formation of Li- compound was also observed. An example of graphene grown on SiC(0001) is displayed by the LEEM image in Fig. 1(a), using a field of view (FOV) of 50 μm. The electron reflectivity curve recorded from the bright grey area and the darker grey area represent 1 ML graphene and a buffer layer (0ML), respectively. The weak lines seen across the sample surface represent steps and illustrate the step bunching that occurs on the SiC substrate during graphene growth. Images recorded directly after Li deposition show significant differences as illustrated in Fig.1(b). The surface appears rougher, grainy, after Li deposition but nevertheless well ordered as confirmed by the ($\sqrt{3}$x$\sqrt{3}$) R30° reconstruction μ-LEED pattern observed on the monolayer area.[26-27] This was suggested to indicate formation of a $C_6Li$ compound[49] or Li adsorption in the hollow sites on graphene. The surface morphology was surprisingly found to deteriorate and not improve after heating at different temperatures. However, after heating the sample at ~260°C, Fig. 1(c), the mosaic pattern can no longer be observed on the 0ML area but wrinkles/cracks along the steps can still be observed on the monolayer terraces. For Na, no intercalation at the interface was observed after deposition when the sample was kept at room temperature. Annealing at about 100°C was found to require activating the penetration process.[36] No intercalation was observed for Rb and Cs.[35]

Calculated values of distance between Li and Na adatoms and free standing single layer graphene is near to the previously calculated values[42] and close to the values calculated by us for the model of epitaxial graphene (Fig. 3a). Increase of the number of alkali adatoms on graphene result insignificant changes in the graphene-alkali distance for both species and enormous increase of the total doping value for the case of lithium (see insets on Fig. 3b). More distant sodium adatoms demonstrate absence of the grown of the doping simultaneously with increase of impurity concentration. For the 100% coverage by lithium (Fig. 2d) total charge transfer from the all adatoms to the graphene is nearly 3 electrons that should correspond with the shift of Fermi energy about 3eV[42] and diminishment of the vacancy formation energy to few eV.[43] Experimental results[26] also report significant (1.2 eV) shift of Fermi energy actually after the penetration process when bigger part of lithium adatoms passivated the outermost Si-C bi-layer of SiC substrate and/or defects in buffer layer and did not dope graphene. In our calculations is also observed significant decay of the energy barrier for the first step of penetration caused by increasing of the concentration of the adatoms. Other source of diminishment of the energy barrier is formation of the uniform environment of re-location or absent of carbon atoms from graphene sheet in the process of penetration (see Fig. 2e).

Reported on the Fig. 3b values of the energy barriers suggest for the possibility of penetration alkali adatoms throughout graphene membrane in the case of 100% coverage. This result is in good agreement with experimentally observed penetration of the Li and Na drop throughout upper layer of epitaxial graphene. Next steps of our survey is the calculation of the energy barriers for the all steps of the penetration process for the both types of graphene (epitaxial and free standing) and both studied types of alkali impurities. Results of the calculations reported on Fig. 3c. From this figure we could conclude that the energy difference between even and odd steps on the intermediate stages of penetration process is about 0.8 eV for lithium and 2 eV for sodium. The values obtained for the first step of lithium penetration (0.3 eV) is smaller than the energy required for water evaporation, defined as the difference between the formation energies of water in the liquid and gaseous phases (43.98 kJ/mol[51] or 0.46 eV). Calculated energy barrier for the first step of sodium penetration (above 0.8 eV) we compared the calculated energy required for the monolayer graphene oxidation (1.35 eV)[52] with the experimental results of graphene monolayer fast oxidation at 200 °C.[53] These values are in the good agreement with the obtained experimental results. From Fig. 3c we could also conclude important difference between epitaxial and free standing graphene. For the last case the process is reversible and migration of adatoms possible in both sides. For the case of epitaxial graphene the placement of all alkali atoms between graphene and buffer layer (Fig. 2g) is more energetically favorable than initial coverage of graphene from above (Fig. 2d). The energy barrier for the migration of first atom from the space between graphene and buffer layer to the surface graphene is about 1.1 eV for Li and 2.2 for Na. This value is evidence for irreversibility of the penetration process in term of energy compare to the free-standing graphene. It also necessary to note that for the case of real epitaxial graphene alkali atoms after penetration throughout graphene sheet will passivate the surface defects in buffer layer and SiC substrate and that should decrease the energy barriers for the intermediate stages of migrations and makes process of penetration more irreversible.

The calculation performed for the rubidium and cesium evidence that the maximal possible coverage by these species is 50% and 25% and the energy barrier for the penetration is about 4 and 6 eV. That is corresponding for the values calculated for the He migration throughout graphene membrane[20] and in agreement with the experimental results of the impenetrability of graphene for helium.[21]

## 4. Conclusions

Calculated values of the energy barriers for the penetration of alkali adatoms throughout graphene membrane by substitution of carbon atoms suggest rather low values of the energy barriers for the formation of temporary defects in carbon layer for in the presence of lithium, two times bigger for sodium and the impossibility of penetration for rubidium and cesium. These results are in agreement with recent experimental findings, i.e. the intercalation studies of alkaline metals on epitaxial graphene grown on SiC(0001) substrates.[26,27, 35,36] However, this intercalation mechanism is significantly different from those reported on the graphite intercalated compounds[54,55] where the inserting of the atoms and molecules between graphene planes required heating of the graphite substrate and is possible for atoms and molecules which are larger than rubidium.

Our finding therefore reveals a different possibility for the penetration of atoms throughout doped graphene at room temperature. In addition to these, it makes possible a fabrication of molecular filters, which require our finding information.[37-41]

**Acknowledgements** DWB thank Korea Institute for Advanced Study for providing computing resources (KIAS Center for Advanced Computation Linux Cluster System) for this work. CV would like to thank supports from the European Science Foundation, EuroGRAPHENE program and the EU project, ConceptGraphene.

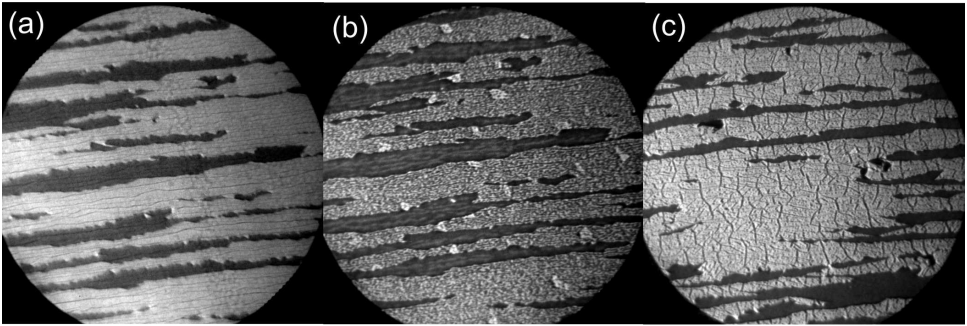

**Figure 1** LEEM images taken at an FOV of 15 µm from a sample with a mixture of 0 and 1 ML graphene areas grown on SiC(0001) (modified from ref [27]) : (a) before deposition and at an electron energy of 0.4 eV, (b) after Li deposition and at electron energy of 4.5 eV and (c) after heating at 260 °C and at an electron energy of 6.8 eV.

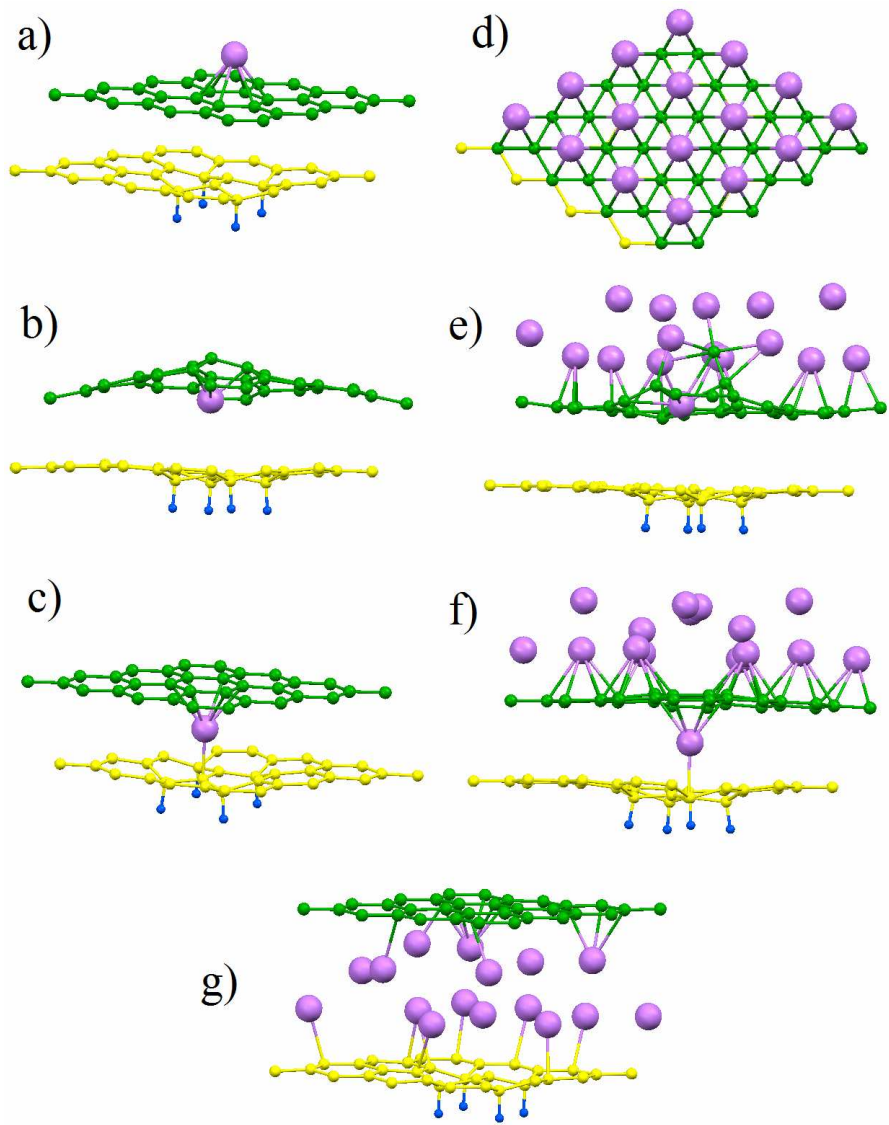

**Figure 2** Optimized atomic structure of initial (a and d), intermediate (b and e), and second (c and f) steps of penetration of Li atom throughout graphene membrane for 12.5 (a - c) and 100% (d – f) coverage of graphene surface by lithium, and final (after penetration of all Li atoms) configuration for the 100% coverage. Lithium atoms are shown by large violet spheres, carbon atom from top graphene layer by small green spheres, and carbon and hydrogen atoms from the underlayer (see description in text) by yellow and blue small spheres respectively.

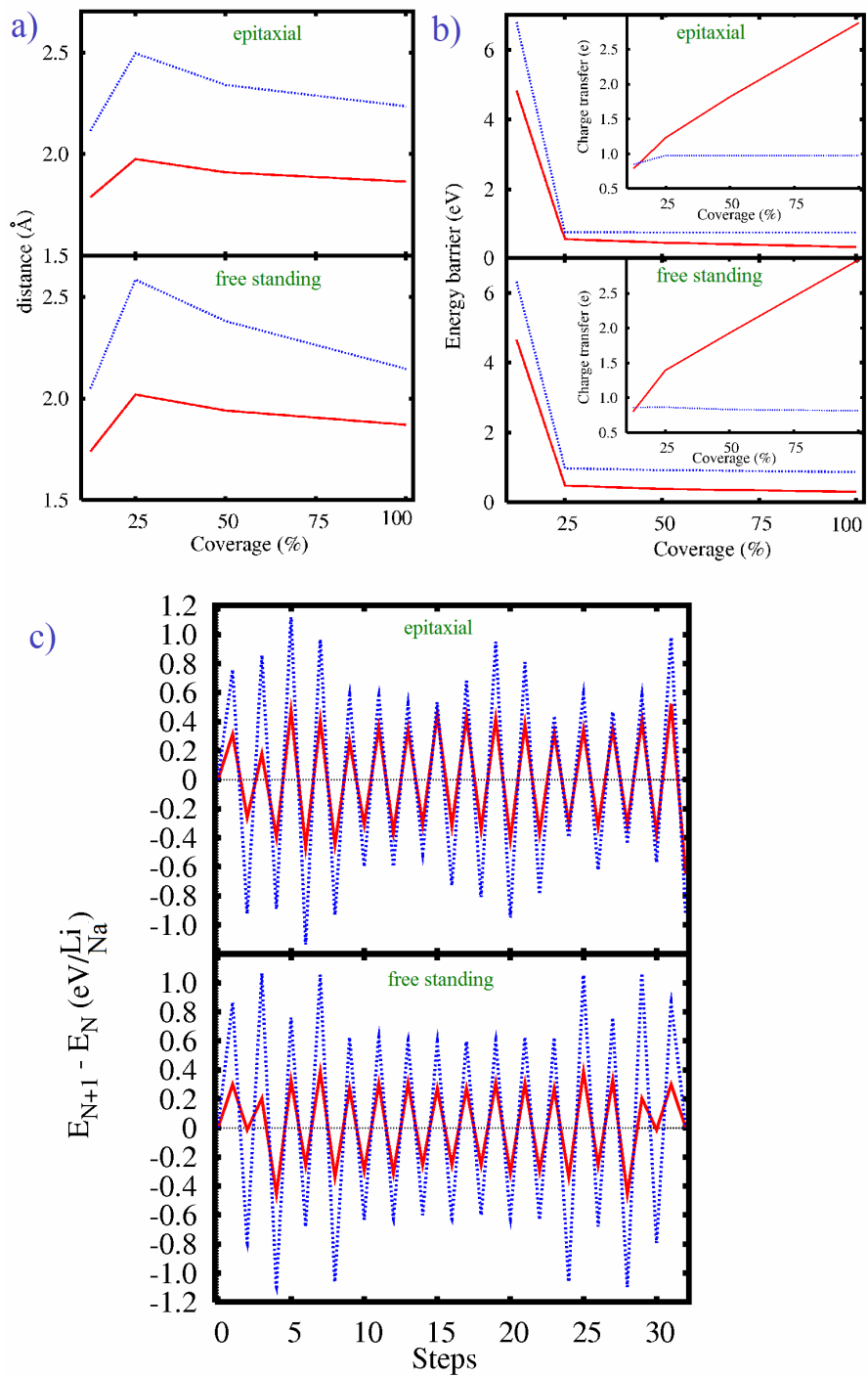

**Figure 3** (a) Distance between graphene flat and alkali adatoms; (b) energy barriers for the first steps of penetration and (on insets) charge transfer (in electrons) as function of coverage. (c) Step-by-step energetics of penetration for the 100% coverage. Results for the lithium are shown by solid red lines and for sodium by dotted blue lines.